\documentclass{aa}
\usepackage{graphicx,natbib}
\bibpunct{(}{)}{;}{a}{}{,}

\def\mnras{MNRAS}

\def\lum{\rm erg~s$^{-1}$}

\begin{document}

\sloppypar

   \title{Low-mass X-ray binaries in the bulge of the Milky Way}

   \author{M.~Revnivtsev \inst{1,2}, A.~Lutovinov \inst{2}, E.~Churazov \inst{1,2}, S.~Sazonov\inst{1,2}, M.~Gilfanov\inst{1,2}, S.~Grebenev\inst{2}, R.~Sunyaev\inst{1,2}}

   \offprints{mikej@mpa-garching.mpg.de}

   \institute{
              Max-Planck-Institute f\"ur Astrophysik,
              Karl-Schwarzschild-Str. 1, D-85740 Garching bei M\"unchen,
              Germany,
     \and
              Space Research Institute, Russian Academy of Sciences,
              Profsoyuznaya 84/32, 117997 Moscow, Russia
            }
  \date{}

        \authorrunning{Revnivtsev M. et al.}
        \titlerunning{LMXBs in the bulge of the Milky Way}

\abstract{We study the population of low-mass X-ray binaries (LMXBs) in the
Galactic bulge using the deep survey of this region by the IBIS
telescope aboard the INTEGRAL observatory. Thanks to the increased
sensitivity with respect to previous surveys of this field, we
succeeded to probe the luminosity function (LF) of LMXBs down to $\sim
7\times10^{34}$ \lum\ in the 17--60~keV energy band. The slope $d\log
N/d\log L=-0.96\pm0.20$ measured in the $10^{35}$--$10^{37}$ 
\lum\ range confirms that the LMXB LF flattens below $L_{\rm x}\la 10^{37}$
\lum\ with respect to higher luminosities. We discuss the origin of
the observed LF flattening. We demonstrate that the spatial
distribution of persistent LMXBs in the Galactic Center/Galactic bulge
region is consistent with a model of stellar mass distribution that
includes the nuclear stellar disk component in the innermost degree of
the Galaxy. The spatial distribution of transient LMXBs
detected in the Galactic Center region indicates an increased fraction
of transient sources in the innermost degree of the Galaxy with respect to 
outer regions.
}

   \maketitle

%

\section{Introduction}

Low-mass X-ray binaries (LMXBs) are binary systems with a compact
object that accretes matter from a low mass optical companion
star filling its Roche lobe \cite[see e.g][]{vdh75}.
Ways of formation and evolution of LMXBs are important and not yet
fully understood questions of X-ray astronomy (see 
discussions of these issues in e.g. \citealt{webbink83,lipunov07,b08}).

The shape of the luminosity function (LF) of LMXBs may provide important
information about their long term evolution. In 
particular, \cite{postnov05} noted that the observed shape of the LMXB
LF (assuming that the X-ray luminosity of the compact object
is directly proportional to the mass transfer rate in the binary
system) depends on the distribution of the masses of optical donor stars
in LMXBs, which is in turn intimately related to the mechanism of
removal of the angular momentum of the binary system. They
suggested that the flattening of the LMXB LF at luminosities
$L_{\rm x}<2\times10^{37}$ \lum\ found in a number of studies
\citep{primini93,gilfanov04,voss06,voss07} might correspond to the
transition from the magnetic stellar wind braking to the
gravitational wave braking mechanism of removal of the binary
system angular momentum \citep{postnov05}. Confirmation of this 
idea would provide additional evidence for the action of gravitational 
braking mechanisms in LMXBs and enable one to select those binary
systems effectively emitting gravitational waves for the next
generation of gravitational wave detectors. It is therefore 
interesting and important to check the presence of the break in the
LMXB LF and to continue the LF to lower luminosities.

Using LF as a tool for studying populations of X-ray sources in nearby
galaxies is only possible with instruments that have good sensitivity
and angular resolution. First attempts of estimating the LF of bright
X-ray binaries were done with the Einstein/HEAO2 observatory \cite[see
e.g][]{speybroeck79,trinchieri91} more than twenty years ago. Studying
LMXB LF in galaxies has now become quite common due to the
unprecedented capabilities of the {\em Chandra} X-ray
Observatory. However, even this powerful observatory does not allow 
one to probe the faint end ($L_{\rm x}<10^{36-37}$ \lum) of the
LMXB LF in the majority of nearby galaxies. This is only feasible for
the nearest galaxies, like M31 \citep{trinchieri91,primini93,voss07}. 

The Milky Way is a unique galaxy where we can detect X-ray objects
with the lowest possible luminosities. However, because of the large
angular size of the Galaxy and the different and often poorly known
distances to Galactic X-ray sources, it is not easy to construct the LF of
LMXBs in the Milky Way. First of all, one needs to cover a large fraction of
the sky. However, focusing X-ray telescopes usually have small fields of
view ($\sim$10~arcmin) and thus cannot cover a considerable part of
the Galaxy within a reasonable time. Even the region of maximal
concentration of Galactic LMXBs -- the Galactic bulge -- is too large
for focusing X-ray telescopes. 

The Galactic Center/Galactic bulge region is most suitable for
studying Galactic LMXBs because the majority of sources are
spatially concentrated and are located at nearly the same distance
from the Sun. \cite{skinner93} reconstructed the surface
density map of X-ray sources in the Galactic bulge region using a
compilation of observational campaigns by different
observatories. The population of Galactic bulge X-ray sources was also
studied with the exceptionally wide field-of-view ART-P coded-mask  
telescope aboard the {\em GRANAT} observatory \citep{grebenev96}.

\cite{grimm02} used data of the All-Sky Monitor (ASM) of the {\em
  RXTE} observatory in the 2--10~keV energy band to construct the
  luminosity function of Galactic LMXBs with luminosities 
$L_{\rm x}\ga 10^{36}$ \lum. Later, \cite{gilfanov04} studied the
populations of LMXBs in the Milky Way and nearby galaxies and
showed that the LMXB LF has a break at a luminosity $L_{\rm
  x}\sim2\times10^{37}$  \lum, below which it significantly 
flattens. Similar flattening of the LMXB LF was found by
  \cite{primini93} for the population of sources in the M31
galaxy. Recent studies of a number of galaxies performed with the {\em 
Chandra} observatory confirmed such a break in the LMXB LF in
the M31 galaxy \citep{kong03,voss07} and indicated the presence of a
break for the Cen A galaxy \citep{voss06}. No breaks, however, were
detected by \cite{kim06} in the LF of sources in NGC3379 and NGC4278.

In order to make a significant improvement in the statistical study
of LMXBs in our Galaxy and independently measure the faint end of the
LMXB LF, one needs a large field-of-view instrument with increased (in
comparison with previous ones) sensitivity. It is also necessary to
use a hard X-ray energy band to avoid contamination by large numbers
of nearby, soft X-ray emitting stars, as was the case with the ROSAT
all-sky survey \cite[e.g.][]{voges99}. This is exactly what is provided by the
coded-mask telescopes of the INTEGRAL observatory \citep{integral}.  
Deep observations of the Galaxy performed with INTEGRAL/IBIS over five
years of its operation have allowed us to increase the sensitivity to
point sources by an order of magnitude in comparison with the
RXTE/ASM survey. This allows us to probe the faint end of
the LF of Galactic LMXBs for the first time.

\section{Sample of sources and data analysis}

To avoid problems with unknown source distances and consequently
unknown source luminosities, we do not use here the whole INTEGRAL survey of
the Milky Way and restrict ourselves to the Galactic
Center/Galactic bulge region where LMXBs are strongly concentrated \cite[see
e.g][]{skinner93,grebenev96,grimm02,lutovinov05}.

The INTEGRAL observatory has spent a lot of time observing the Galactic Center
(Galactic bulge) region over the first five years of its operations \cite[see
e.g][]{revnivtsev04,kuulkers07}. Among the INTEGRAL instruments, the IBIS
telescope best meets the requirements of the present study: it has good
sensitivity for detecting point sources and a wide field of
view ($\sim28^{\circ}\times28^{\circ}$). Due to the hard energy band
of IBIS ($>17$ keV), it is less sensitive to sources that have high
luminosities in the standard 2--10 keV energy band (since such sources
have very soft spectra that do not continue to hard X-rays), but it is
very effective in detecting low-luminosity objects (which typically have hard
spectra). The sensitivity achieved by now with respect to
point sources in the Galactic Center region is typically 0.5 mCrab in
the 17--60~keV energy band \citep[see e.g.][]{krivonos07}. 

For the present study we selected all sources detected by INTEGRAL in the
Galactic Center (Galactic bulge) region listed in the catalog of
\cite{krivonos07} and added INTEGRAL/IBIS data that became available since the
publication of this catalog. These additional data allowed us to increase the
accuracy of source flux measurements and to add one newly detected source, IGR
J17586--2129, to the sample. 

We only considered those sources located within the elliptical region around
the Galactic Center with axes $|l|<10.7^\circ, |b|<5.1^\circ$ 
and having fluxes higher than 0.64 mCrab in the 17--60~keV energy
band, which typically ensures the source detection with statistical
significance more than 8$\sigma$. This conservative detection
threshold prevents any problems with possible systematical
uncertainties in the image reconstruction and with variations of the
sensitivity over the region of our interest. 

Due to the anticipated concentration of LMXBs in the Galactic bulge,
we assume that all sources that are not known to be high-mass 
X-ray binaries, cataclysmic variables or extragalactic sources, are
LMXBs. It is possible that a detailed study of some faint sources
(typically with luminosities $L<10^{35}$ \lum) from our sample  
will reveal their non-LMXB nature. Therefore, we are probably somewhat
overestimating the faint end of the LMXB LF.

From the preliminary source list we have filtered out:

\begin{enumerate}
\item {\bf HMXBs}. High-mass X-ray binaries (HMXBs) are 
young objects, which in our Galaxy are concentrated to the regions of recent
star formation -- the spiral arms. The spatial distribution
of Galactic HMXBs has been studied by \cite{grimm02} in
soft X-rays and by \cite{lutovinov05,lutovinov07} in hard X-rays
($>20$ keV) with the INTEGRAL observatory. Due to the low
star-formation rate in the Galactic bulge we can anticipate 
that HMXBs should not be a dominant population of bright X-ray
sources there. However, we might detect some HMXBs located in the
Galactic disk, projected on the bulge in the sky plane. We excluded
from our analysis all sources that are known or supposed to be HMXBs,
for example so-called supergiant fast X-ray transients (SFXTs) and
X-ray pulsars. These sources include EXO 1722$-$363, IGR J17407$-$2808,
XTE J1743$-$363, IGR 
J17544$-$2619, IGR J17391$-$3021 (XTE J1739$-$302), AX J1749.1$-$2733, AX
J1749.2$-$2725 and IGR J18027-2016. Note that most of them have a
transient nature. 
 
\begin{table}[htb]
\caption{\scriptsize List of known or likely LMXB sources (persistent and 
transients) in the Galactic bulge region detected by INTEGRAL on the
time-average map. The 1 mCrab flux in the 17--60 keV energy band for
a Crab-like spectrum corresponds to an energy flux of
$\sim1.4\times10^{-11}$ \lum ~cm$^{-2}$ and to a luminosity
$1.1\times10^{35}$ \lum\ for a distance of 8 kpc. Fluxes 
in the 2--10 keV energy band reported for some sources in the last
column were obtained from RXTE/ASM measurements or adopted from the
literature if a reference is given. The hard X-ray fluxes of the
sources SLX 1744$-$299 and SLX 1744$-$300, which are not resolvable with
INTEGRAL/IBIS, were calculated from their measured summed flux by
assuming their flux ratio to be 2:1 \citep{sidoli99}} 
\label{list} \tabcolsep=3mm
\begin{tabular}{l|c|c|c}
\hline
\#&Name&{Flux, mCrab} & Flux, mCrab\\
&&{17-60 keV} & 2-10 keV \\
\hline
1&GRS 1758-258  &      59.07    $\pm$  0.07 &     25\\
2&GX 1+4        &      54.14    $\pm$  0.08 &     11 \\
3&GX 5-1        &      48.14    $\pm$  0.07 &     972 \\
4&GX 354-0      &      37.57    $\pm$  0.08 &     99 \\
5&1E 1740.7-2942 &     31.79    $\pm$  0.07 &     10 \\
6&4U 1724-30    &      17.14    $\pm$  0.07 &     25 \\
7&A1742-294     &      13.71    $\pm$  0.07 &     18[1] \\
8&GX9+1         &      16.50    $\pm$  0.09 &     532 \\
9&SLX 1735-269  &      11.21   $\pm$     0.07 &     14\\
10&GX3+1        &       9.93   $\pm$     0.07 &     300 \\
11&SLX1744-299 &        5.47   $\pm$     0.07 &     9.0[1]\\
12&1E1742.8-2853   &       6.29 $\pm$       0.07 &       \\
13&1E1743.1-2843   &       5.20 $\pm$       0.07 &      8.1[2] \\
14&SLX 1737-282    &       3.94 $\pm$       0.07 &      5.4[3] \\
15&SLX1744-300     &       2.73 $\pm$       0.07 &      5.1[1]\\
16&IGR J17254-3257 &       1.72 $\pm$       0.08 &      3.5 \\
17&IGR J17475-2253 &       1.09 $\pm$       0.08 &      \\
18&IGR J17353-3539 &       0.88 $\pm$       0.08 &       \\
19&IGR J17505-2644 &       0.81 $\pm$       0.07 &       \\
20&IGR J17585-3057 &       0.81 $\pm$       0.07 &       \\
21&IGR J17586-2129 &       0.73 $\pm$       0.08 &       \\
22&IGR J17448-3231 &       0.70 $\pm$       0.07 &       \\
\hline
\multicolumn{4}{c}{Transients detected on the averaged map} \\
\hline
1&GRS 1741.9-2853   &       3.17  $\pm$      0.07 &  \\
2&SAX J1747.0-2853  &       3.30  $\pm$      0.07 &  \\
3&KS 1741-293       &       4.95  $\pm$      0.07 &  \\
4&XTE J1739-285     &       2.21  $\pm$      0.07 &  \\
5&GRS 1747-313      &       1.53  $\pm$      0.07 &  \\
6&IGR J17464-3213   &      22.07  $\pm$      0.07 &  \\
7&SLX 1746-331      &       0.69  $\pm$      0.07 &  \\
8&IGR J17353-3257   &       1.12  $\pm$      0.07 &  \\
9&MXB 1730-33       &       4.20  $\pm$      0.08 &  \\
10&IGR J17331-2406   &       1.01 $\pm$       0.08 &  \\
11&XTE J1720-318     &       1.73 $\pm$       0.08 &  \\
12&IGR J17597-2201   &       4.62 $\pm$       0.08 &  \\
13&XTE J1817-330     &       7.01 $\pm$       0.09 &  \\
14&4U 1746-37        &       2.77 $\pm$       0.09 &  \\
15&4U 1705-32        &       2.30 $\pm$       0.09 &  \\
16&IGR J17091-3624   &       4.75 $\pm$       0.10 &  \\
\hline
\end{tabular}
\begin{list}{}
\item  [1] - \cite{sidoli99}, [2] - Porquet et al. 2003, [3] - in't Zand et al. 2002
\end{list}
\end{table}

\item {\bf Extragalactic}. We also excluded extragalactic sources.
Within the region of our study there are only two known
extragalactic sources -- active galactic nuclei GRS 1734$-$292 and IGR
J17488$-$3253. According to the all-sky averaged AGN number-flux function
in hard X-rays \citep{krivonos07} we should detect $\sim2$
AGNs above the adopted flux limit of $0.64$ mCrab within the area of
our study ($\sim171$~sq.~deg). This indicates that most likely we
already know of all the extragalactic sources with fluxes of $>0.64$ 
mCrab within the considered field. The probability that there are 
additional 2 or more unidentified extragalactic sources in this
region is $\sim$0.14 and we do not expect unknown extragalactic sources
to considerably contribute to the measured number-flux function of
sources within the bulge region.

\item {\bf Other sources} We removed cataclysmic variables, detected
in the considered region. We removed from our sample the source
IGR J17402$-$3656, which is associated with the nearby open star cluster
NGC 6400 and may be composed of multiple point sources; the source IGR
J17456$-$2901, which is similarly associated with the collective
emission of the nuclear stellar cluster
\citep{belanger06,krivonos07a}; and the source IGR J17475$-$2822
representing the emission of the molecular cloud Sgr B2 
\citep{revnivtsev04b}. 
\end{enumerate}

\subsection{Transients}

A significant fraction of Galactic LMXBs are transients. During
outbursts their luminosities can change by more than 6--7 orders
of magnitude \cite[e.g.][]{tanaka96}. The "on" state of bright X-ray
transients typically lasts several weeks, but due to
their extreme brightness at that time they may still be detected 
on a sky map averaged over several years even if they were
undetectable most of the time. 

It is usually considered that the observed X-ray brightness variations
of LMXB transients are caused by a developed mass transfer instability
in the accretion disk \cite[e.g.][]{meyer84,lasota01}. In this case, the X-ray 
luminosity observed from such a source at any given time does not
provide us information about the global mass transfer rate in
the binary system, which is of interest for us. The global mass
transfer rate could be determined by averaging the X-ray luminosity of
a transient source over a period that is much longer than the
accretion disk mass accumulation time; however, this time scale might
be too long \cite[e.g.][]{chen97} to be testable by available X-ray
observations. The inclusion of transient sources with their fluxes
averaged over some arbitrary period (not related to any physically
motivated time scale) might distort the true distribution of mass
transfer rates, in which we are interested. 

Therefore, in our analysis we treated the transient sources in two
ways: 1) we included all sources detected at the 5-year averaged
INTEGRAL/IBIS map of the Galactic Center, and 2) we filtered out all
transient sources from our sample. In the latter case, we adopted the
following selection criterium for separating transient sources from
persistent ones: a transient should have the ratio of maximal 3-day
averaged flux to the all-time mean flux of more than 5.0. We note that
transient nature is difficult to ascertain for weaker sources
because they become detectable only after accumulating a large part of
the total exposure time. Therefore, our resulting sample of transient
sources is likely significantly incomplete at low fluxes.

\begin{figure}
\includegraphics[width=\columnwidth]{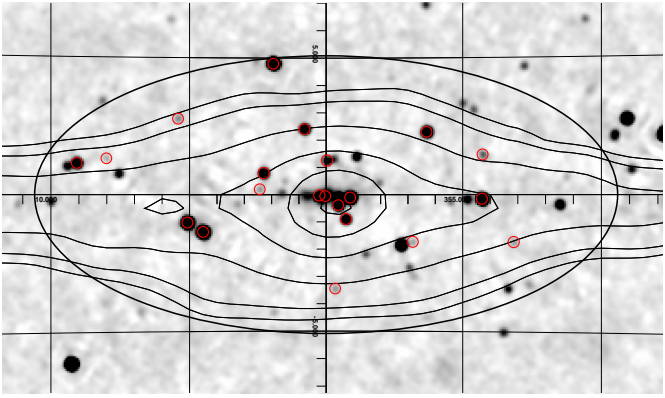}

\caption{Map of the Galactic Center region in the 17--60 keV energy
band obtained by INTEGRAL/IBIS. The ellipse encloses the region of
our study. Sources used in the analysis of the LMXB LF are marked by circles.
Contours are isophotes of the 4.9$\mu m$ surface brightness of
the Galaxy (COBE/DIRBE) demonstrating the bulge/disk structure of
the inner Galaxy.}  \label{image}

\end{figure}

The final list of accepted sources is presented in Table \ref{list}.
The map of the Galactic bulge region with the sources from Table
\ref{list} is presented in Fig.~\ref{image}. 

\subsection{Distances to the sources}

The distances to the majority of our sources are unknown. However,
assuming that LMXBs are distributed in the Galactic bulge similarly to
ordinary stars, we can calculate the density distribution of LMXBs in the bulge
and predict the spread of values of the flux to luminosity conversion
coefficient for our sources. This determines the typical systematic error we
are making in estimating source luminosities by assuming that all
sources are at the same, 8 kpc, distance from the Sun. 

\begin{figure}
\includegraphics[width=\columnwidth]{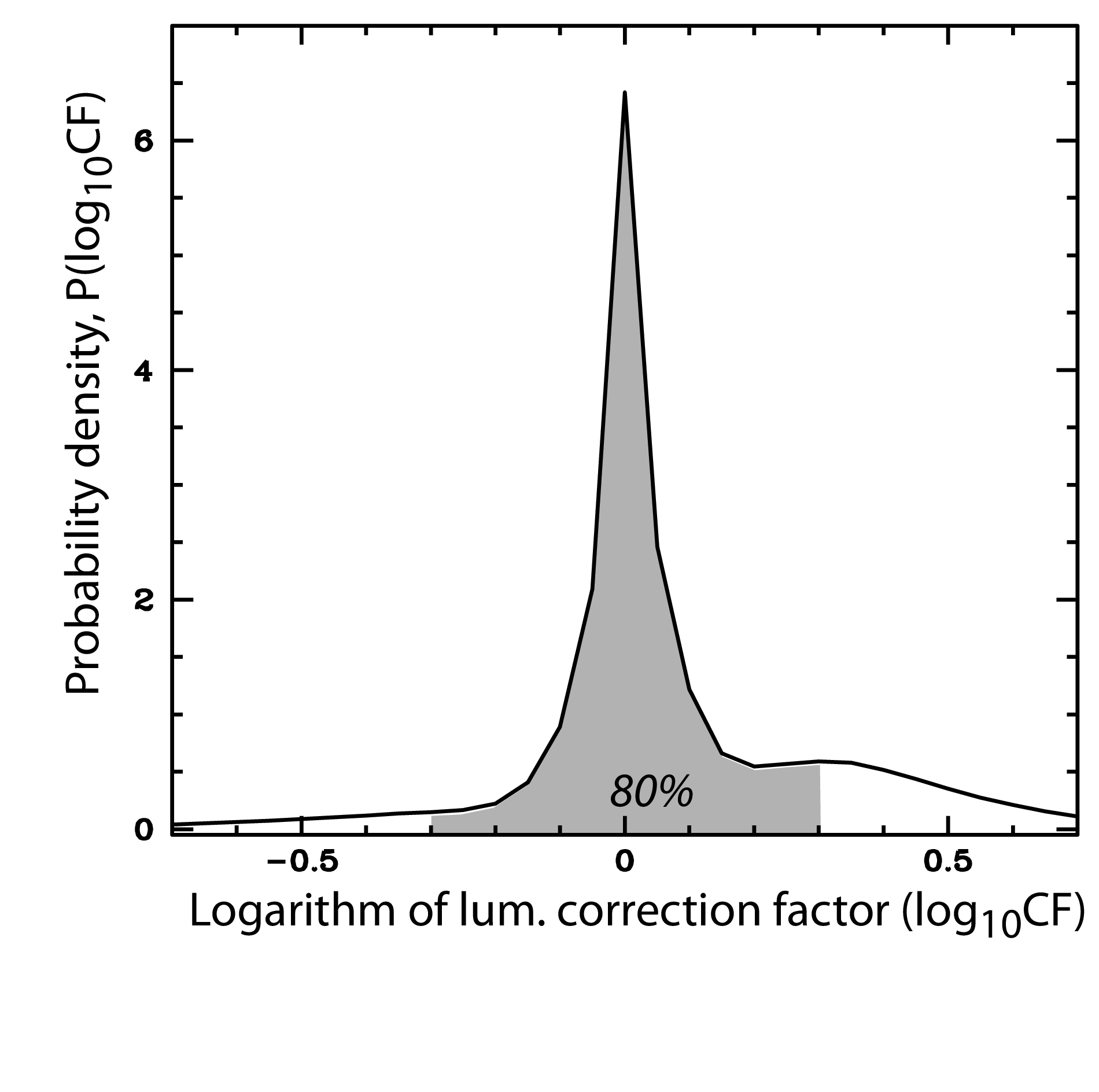}

\caption{Probability density of the luminosity correction factor
  (the probability density is normalized so that $\int P(\log_{10}CF)d(\log_{10}CF)=1$). Eighty
per cent of sources in the studied region of the Galaxy are expected to fall
  into the interval shown in gray.}
\label{prob}

\end{figure}

To this end, we adopted a mass model of the Galaxy consisting of the
Galactic bulge, Galactic disk and nuclear stellar disk
\cite[e.g.][]{launhardt02}. We did not consider the nuclear stellar
cluster, which is important in the innermost $\sim10^\prime$ of the Galaxy,
because its size is smaller than the angular resolution of INTEGRAL/IBIS.

The stellar density in the bulge was assumed to follow model G3
from \cite{dwek95}. We adopted the mass of the bulge to be
$1.3\times10^{10} M_\odot$ \citep{dwek95}. 

The stellar density in the Galactic disk:
$$
\rho_{\rm disk}(r,z)=\rho_{\rm 0,disk} \exp \left[ -\left( {r_m\over{r}} \right)^3 - \left( {r\over{r_{\rm disk}}}\right) - \left({|z|\over{z_{\rm disk}}}\right) \right],
$$
where $r_{\rm disk}=2.2$ kpc, $r_{\rm m}=2.5$ kpc, $z_{\rm
disk}=130$ pc \citep{revnivtsev07}. The total mass of the Galactic stellar
disk was taken to be $2.5\times10^{10} M_\odot$.

The nuclear stellar disk (NSD) was assumed to have the density distribution
($r$ and $z$ are measured in parsecs)
$$
\rho_{\rm NSD}=\rho_d r^{-\alpha}e^{-|z|/z_d},
$$
where $z_d=45$ pc. At $r<120$ pc the slope $\alpha=0.1$ and the
normalization constant $\rho_d=300 M_\odot$/pc$^3$. At $120$ pc
$<r<220$ pc, $\alpha=3.5$, and at $r>220$ pc, $\alpha=10$
\citep{launhardt02}, and we adjust the constant $\rho_d$ so that the
density distribution is continuous. The total adopted mass of
the NSD is $1.4\times10^9 M_\odot$. In reality this quantity is
uncertain by some 50\% \citep{launhardt02}.

Based on this model, we determined the distribution of stars over the
distance from the Sun in the region of our study
(ellipse $|l|<10.7^\circ, |b|<5.1^\circ$). The ratio $(D/8~{\rm
  kpc})^2$ defines the luminosity correction factor $CF$. The
probability density of this correction factor can be calculated from
the probability density $P(D)$ of the source distance, which can
be found by integrating the distribution of stellar mass
$\rho_\ast(l,b,D)$ in a given direction $(l,b)$ over the solid angle $\Omega$
of our study:

$$
P(D)={ \int_{\Omega} \rho_{\ast}(l,b,D) D^2  d\Omega \over{ \int_{\Omega}\int_{\rm D=0}^{D=\infty} \rho_{\ast}(l,b,D) D^2  d\Omega}dD}.
$$

The probability density of $CF$ is presented in Fig.~\ref{prob}. It
turns out that for 80\% of sources $\log_{10}{CF}<0.3$. Therefore,
assuming that all bulge sources are located at the same distance
(8~kpc) from the Sun we can predict the luminosities of most
sources to within a factor 2, this uncertainty being comparable to expected
variations in source fluxes.

\subsection{Spectral corrections}

Previously, for constructing the luminosity function of LMXBs in galaxies one
usually used instruments operating in the standard X-ray energy band,
2--10 keV, or similar bands \citep{grimm02,gilfanov04}. The
INTEGRAL/IBIS sky survey is done in the hard X-ray band, 17--60
keV. Therefore, to compare the LMXB LF derived here with those reported before we should take into account the spectral shapes of
LMXBs in the broad energy range 1--100 keV. 

\begin{figure}
\includegraphics[width=\columnwidth, bb=0 100 600 715,clip,angle=-90]{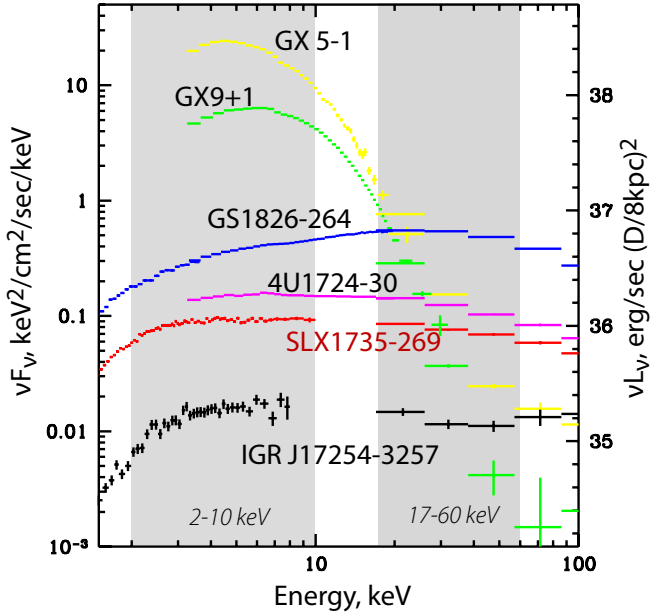}
\caption{Typical broadband spectra of LMXBs. The right axis shows the
logarithm of the luminosity, assuming a distance of 8 kpc to all
sources.} \label{spectra}
\end{figure}

It is well-known that the brightest LMXBs emit most of their
luminosity in the standard X-ray band ($\sim$2--10~keV), while less
luminous objects emit similar amounts of energy in the standard and hard X-ray
($\sim$10--100~keV) bands. Bright LMXBs usually have spectra
with an exponential cutoff at energies of $\sim$6--10~keV
\cite[e.g.][]{rappaport69,toor70,mitsuda84,pavlinsky94} and
essentially do not continue into the hard X-ray band
\cite[e.g.][]{gilfanov93}. In contrast, dim LMXBs are characterized by
a power-law spectral shape between 2 and 60 keV
\citep{mitsuda89,grebenev91,barret91,zhang96,barret00}. We can
therefore expect that the LMXB LF constructed in the 17--60 keV energy
band should strongly differ from that derived in the standard X-ray band.

To demonstrate this, we constructed the broadband spectra of
Galactic bulge sources with different luminosities using all
available instruments -- INTEGRAL/IBIS, RXTE/PCA, Swift/XRT,
ASCA/GIS. This allowed us to cover the wide energy range 1--100 keV (see
Fig.~\ref{spectra}). The LMXB GS 1826$-$24 formally lies outside the
elliptical region of our study, but since its distance from the Sun is
similar to those of bulge sources \cite[e.g.][]{zand99} and its
broadband spectrum has very high statistics this spectrum is also shown in the
figure. 

One can see dramatic changes in the broadband behavior of source spectra at a 
threshold luminosity of $\sim\textrm{few}\times10^{37}$ \lum. At high
luminosities, most of the energy is emitted below 17~keV.  At low
luminosities, the spectral shape in the 2--60 keV band is close to a
power law with photon index $\Gamma\sim$1.9--2 and similar fluxes are
emitted at 2--10 keV and 17--60 keV. 

The time-averaged fluxes of most of the brightest sources in
the 2--10 keV band are directly measured by the All-Sky Monitor
aboard RXTE and we present these measurements in Table \ref{list}.
Based on these data and Fig.~\ref{spectra}, we adopted the following
spectral correction factors for the subsequent analysis:

\begin{displaymath}
{L_{\rm 2-10~keV}\over{L_{\rm 17-60~keV}}} = \left\{ \begin{array}{cl}
40& \textrm{if $L_{\rm 2-10~keV}>2\times10^{37}$ \lum}\\
1.25& \textrm{if $L_{\rm 2-10~keV}<2\times10^{37}$ \lum}\\
\end{array} \right.
\end{displaymath}

In comparing below the LMXB LFs in the 2--10~keV and
17--60~keV energy bands, we mostly used direct RXTE/ASM measurements to
determine the 2--10~keV luminosities for sources with $L_{\rm
2-10~keV}\ga 10^{36}$ \lum, and assumed that those sources with
luminosities below this RXTE/ASM completeness limit are in the hard
spectral state.

It is important to note that the spectral correction factors
introduced above may not be applicable to rare, extremely luminous
LMXBs such as the Galactic source GRS 1915+105, which is more powerful that
any source in our Galactic bulge sample and is apparently in a highly unusual
spectral state.

\section{Results}

\subsection{Luminosity function of faint LMXBs}

\begin{figure}
\includegraphics[width=\columnwidth]{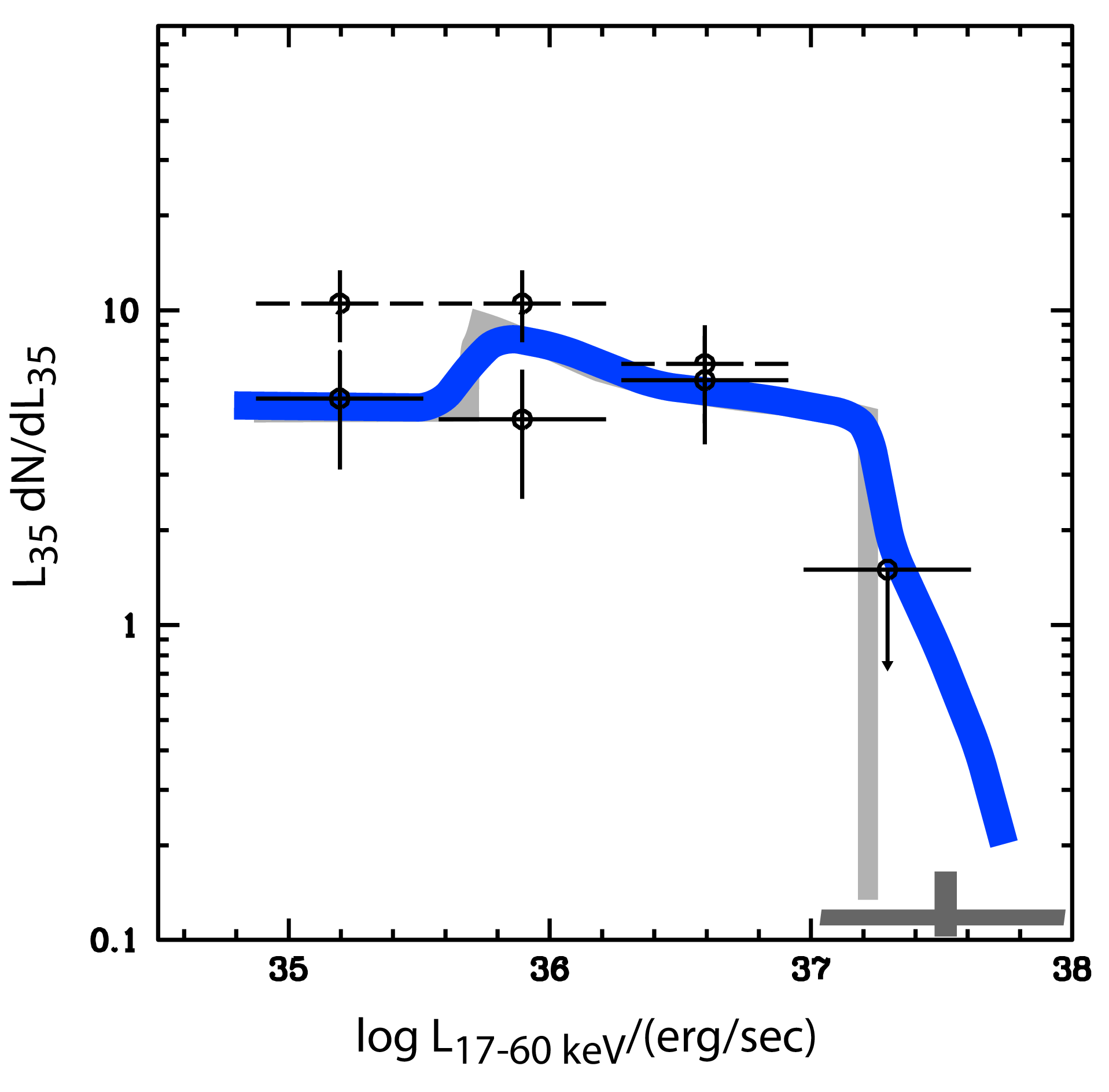}

\caption{Luminosity function of LMXBs detected by INTEGRAL/IBIS in the
Galactic bulge region. The LF of persistent sources is shown by solid
crosses, and that of all sources (including transients) by dashed
crosses. The thick gray cross represents an estimate of the number
of LMXBs with extremely high luminosities based on the one
such source in the Galaxy, GRS 1915+105 ($L_{\rm 17-60~keV}\sim 5\times10^{37}$
\lum). This estimate takes into account that the bulge region covered
by the present study contains only 16\% of the total stellar
mass of the Galaxy. The thick gray line is the analytic model of the
LMXB LF from \cite{gilfanov04} normalized to $1.65\times10^{10}
M_\odot$ -- the stellar mass within the considered region, 
recalculated into the 17--60 keV energy band using the introduced 
spectral correction factors and multiplied by an additional factor of
0.4. The thick blue line is the same LF model convolved with the
distribution of luminosity correction factor, which is important for
sources in our sample due to unknown values of exact distances
(see Fig.\ref{prob})} 
  \label{lf}

\end{figure}

The luminosity functions of all and only persistent LMXBs detected by
INTEGRAL/IBIS in the Galactic bulge region are shown in
Fig.~\ref{lf}. These LFs are defined so as to give the number of sources
within the studied region containing $1.65\times10^{10} M_\odot$ of
stars.

It can be seen that the LF of weak LMXBs ($L_{\rm
 17-60~keV}<10^{36.5}$ \lum) is relatively flat. The LF of all LMXBs
 can be approximated by the function $L_{\rm 35}dN/dL_{\rm 35}=
 (12\pm4)\, L_{\rm 35}^{-(0.13\pm0.13)}$ in the range $10^{34.7}$
 \lum\ $<L_{\rm 17-60~keV}< 10^{36.5}$ \lum, and the LF of peristent
 sources can be approximated by the function 
$L_{\rm 35}dN/dL_{\rm 35}= (4.7\pm2.2) L_{\rm 35}^{(0.04\pm0.2)}$. Here,
$L_{\rm 35}-$ is the hard X-ray (17--60 keV) luminosity in units of
$10^{35}$ \lum. 

There are no objects with $L_{\rm 17-60~keV}>10^{37}$ erg~s$^{-1}$ in
our sample. Consequently, there is an abrupt drop in 
the number density of very bright sources as can be seen in
Fig.~\ref{lf}.  In fact, there is only one non-transient (but strongly
variable) source in the whole Milky Way that has a higher
luminosity: GRS 1915+105 with $L_{\rm 17-60~keV}\sim 5\times10^{37}$
\lum\ (this value was calculated from the time averaged flux of the
source of 261 mCrab in the energy band 17--60 keV from \citealt{krivonos07},   
assuming a 11 kpc source distance, \citealt{harlaftis04}).
Since such ultraluminous sources can be easily detected with INTEGRAL
throughout the Galaxy, we can estimate their expected number in the
Galactic bulge region studied here by taking into account that this
region contains 16\% of the total stellar mass of the Galaxy. This
estimate is shown in Fig.~\ref{lf} by the thick gray cross.

In order to compare the 17--60~keV LF derived here with the
analytic approximation of the 2--10~keV LF presented by \cite{gilfanov04}, we
recalculated the last one into the 17--60 keV energy band using the
spectral correction factors described above. This predicted LF, rescaled 
to the stellar mass contained in the considered region 
($1.65\times 10^{10} M_\odot$), is shown in Fig.~\ref{lf} by the thick 
gray line. It was additionally multiplied by a factor of 0.4 in order 
to match the normalization of the LF measured with INTEGRAL in the 
Galactic bulge. We note that a similar indication of underabundance of 
LMXBs in the Milky Way as compared to other galaxies was
previousely reported by \cite{gilfanov04}.

As is apparent from Fig.~\ref{lf}, transient sources provide a
significant contribution to the LMXB LF. If we describe the LF by the function 
$L_{\rm 35}dN/dL_{\rm 35}=const$ in the luminosity range
$10^{34.7}$--$10^{36.5}$ \lum, the normalization increases by a factor
of $\sim1.7$ after adding the transient sources.

\begin{figure}
\includegraphics[width=\columnwidth]{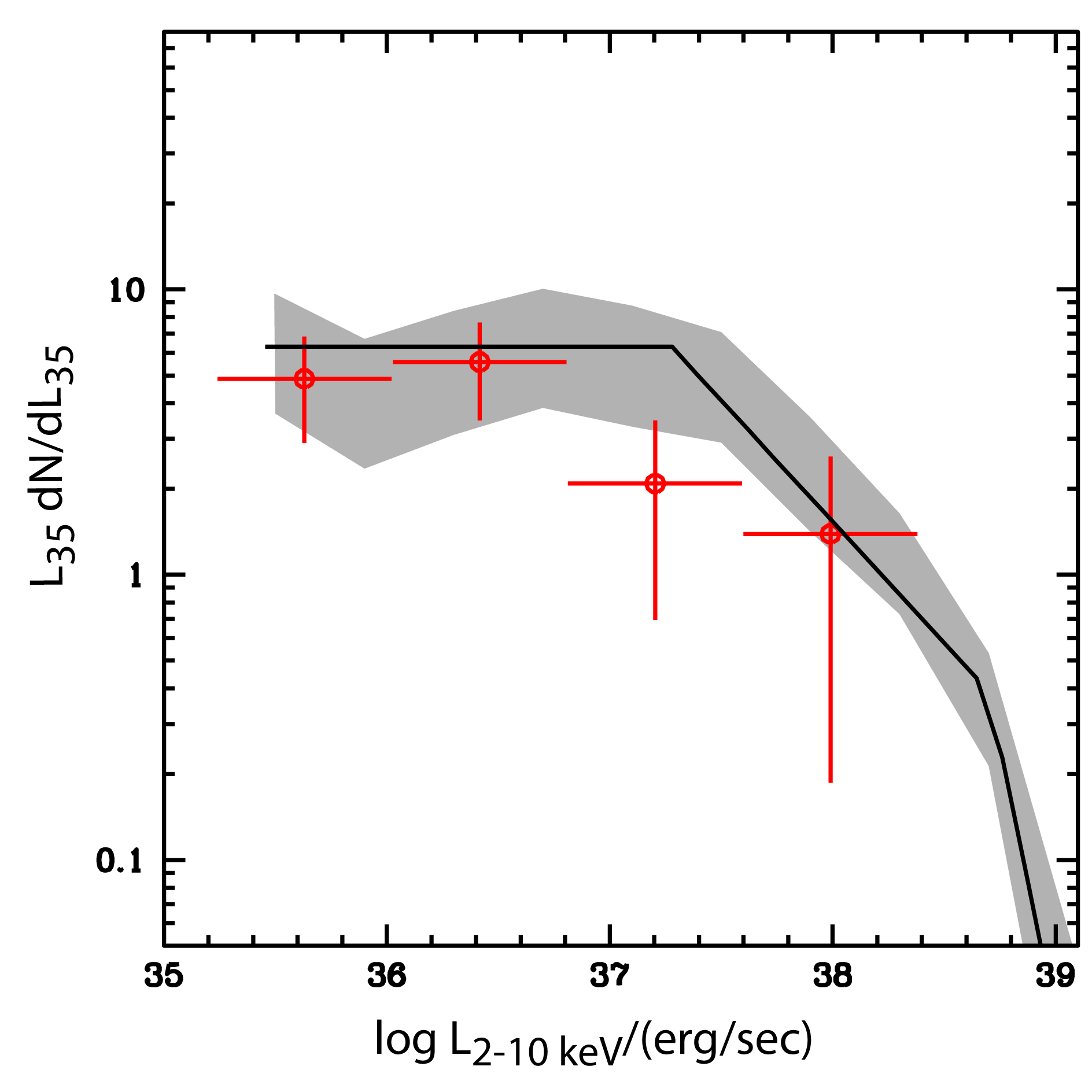}

\caption{Luminosity function of persistent sources in the 2--10 keV
energy band. The solid line shows the analytic form of the LMXB LF
from \cite{gilfanov04} rescaled to $1.65\times10^{10} 
M_\odot$ -- the stellar mass within the region studied here, and multiplied 
by 0.4. The shaded area represents the spread of LF between different
galaxies \citep{gilfanov04}} \label{lfs}

\end{figure}

In Fig.~\ref{lfs} we present the luminosity function of persistent
LMXBs in the standard X-ray band (2--10 keV), which we constructed
using direct RXTE/ASM source flux measurements for brighter sources
(see Table \ref{list}) and our assumed spectral correction factors for
weaker sources. 
  
We should note that, despite the low star-formation rate in the
Galactic bulge, some bulge HMXBs may be present in our sample. This
can be especially important at the lowest luminosities because the LF
of HMXBs is much steeper than that of LMXBs below $\sim2\times10^{37}$
\lum \cite[see e.g.][]{grimm02,gilfanov04}. For example, even for a
star formation rate as low as $\dot{M_\ast}=0.03M_\odot/{\rm 
year}$, the number of faint HMXBs will be comparable to that
of all similarly faint sources in the region of our study. Therefore, 
the true number of faint LMXBs will be even lower than suggested by
the derived luminosity function.

Summarizing all of the above, we can conclude that irrespective of the
inclusion of transient sources the shape of the LF of 
Galactic bulge LMXBs is broadly compatible with the LMXB LF averaged
over nearby galaxies \citep{gilfanov04}, which exhibits a significant
flattening at luminosities $L_{\rm 2-10~keV}<2\times10^{37}$ \lum. 

\subsection{Possible interpretations of the break in the LF}

The flattening of the LMXB LF at luminosities $L_{\rm x}\sim 10^{37}$ \lum\ 
could be due to various reasons: i) a break in the distribution over
mass transfer rates from the companion star, ii) a change of the correspondence
between the mass transfer rate in the binary system and the mass
accretion rate onto the compact object, possibly associated with the
spectral state transition. We now briefly comment on each of these
possibilities.

\subsubsection{State transition}

It is somehow surprising that the break in the LMXB LF occurs
almost exactly at the (broadband) luminosity at which the sources
experience a spectral state transition. At high luminosities, practically all
sources are in the so-called soft/high spectral state and emit most of
their luminosity in the optically thick regime. At low
luminosities, the sources are in the so-called low/hard spectral
state, emit in the optically thin regime and have hard spectra.

It is usually assumed that in the soft/high spectral state the X-ray
luminosity is directly proportional to the mass accretion rate in
the binary system \cite[e.g.][]{ss73}. The dependence of X-ray luminosity
on mass accretion rate in the low/hard state is not clear. In the
framework of the radiatively inefficient accretion flow 
models \citep{ichimaru77,torbett84,narayan94,blandford99}, the
accretion flow X-ray luminosity in the hard state may be a non-linear
function of the mass accretion rate at the outer edge of the accretion
flow: $L_{\rm x}\propto \dot{M}^\beta$, where $\beta>1$. Some
observational data favor such a dependence \cite[e.g.][]{gallo06}.

In this case, even if the distribution of mass transfer rates in
binary systems has no peculiarities at the rates corresponding to
X-ray luminosities $\sim \textrm{few}\times 10^{37}$ \lum, one
would still observe a break in the X-ray luminosity function due
to the change in the $L_{\rm x}-\dot{M}$ relation. If the underlying
distribution of mass transfer rates has the slope $d \log N /d
\log \dot{M}=-\alpha$ and $L_{\rm x}\propto\dot{M}^{\beta}$, then
the resulting slope of the LF will be
$\gamma=(\alpha+\beta-1)/\beta$. Assuming that the true mass
transfer rate distribution has the slope $\alpha=1.8$--2.0, as seen
at luminosities $\log L_{\rm x}\sim$37.5--38.5 
\citep{gilfanov04}, and that $\beta=3.3$ \citep{beckert02},  the
slope of the LF will be $\gamma\sim1.2-1.3$. Such a flattening of the
LF is significant but not strong enough to explain the observed
behavior of the LF.

The majority of the considered X-ray sources are likely accreting
neutron stars. They have solid surfaces and the simple advection dominated
model, in which the accreted mass flows under the black hole horizon without
creation of the X-ray emission, is not directly applicable to them.
However, some other mechanisms of reducing the X-ray emissivity of
the accretion system might be considered \cite[see
  e.g.][]{torbett84,blandford99}.

\subsubsection{Change of the binary braking mechanism}

Following \cite{postnov05}, the break in the LMXB LF can be
interpreted as the luminosity at which the mass transfer rate in the
binary systems starts to driven by the gravitational braking
mechanism rather than by magnetic stellar wind braking.

If this is indeed the case, we can anticipate that the majority of such
faint LMXBs should have very small masses of optical companions
($M_{\rm opt}\la0.2-0.4M_\odot$, see \citealt{postnov05}) and short 
orbital periods ($<2-3$ hours), and be very faint in the optical and
near-infrared spectral bands: the observable brightness of sources
could be as low as $m_R>20$ even without any interstellar extinction
(see estimates of the optical brightness of compact binary systems in
\citealt{vp94}). These predictions can be directly checked when
accurate localizations of such X-ray sources will be achieved with the
help of X-ray telescopes. Positional accuracy of the order of few
arcseconds or less is necessary in this case, because the spatial
density of such faint optical objects ($m_R>20-22$) is very high 
in the direction of the Galactic bulge.

\subsection{Distribution of LMXBs in the Galactic Center region}

\subsubsection{Persistent sources}

\begin{figure}
\includegraphics[height=\columnwidth]{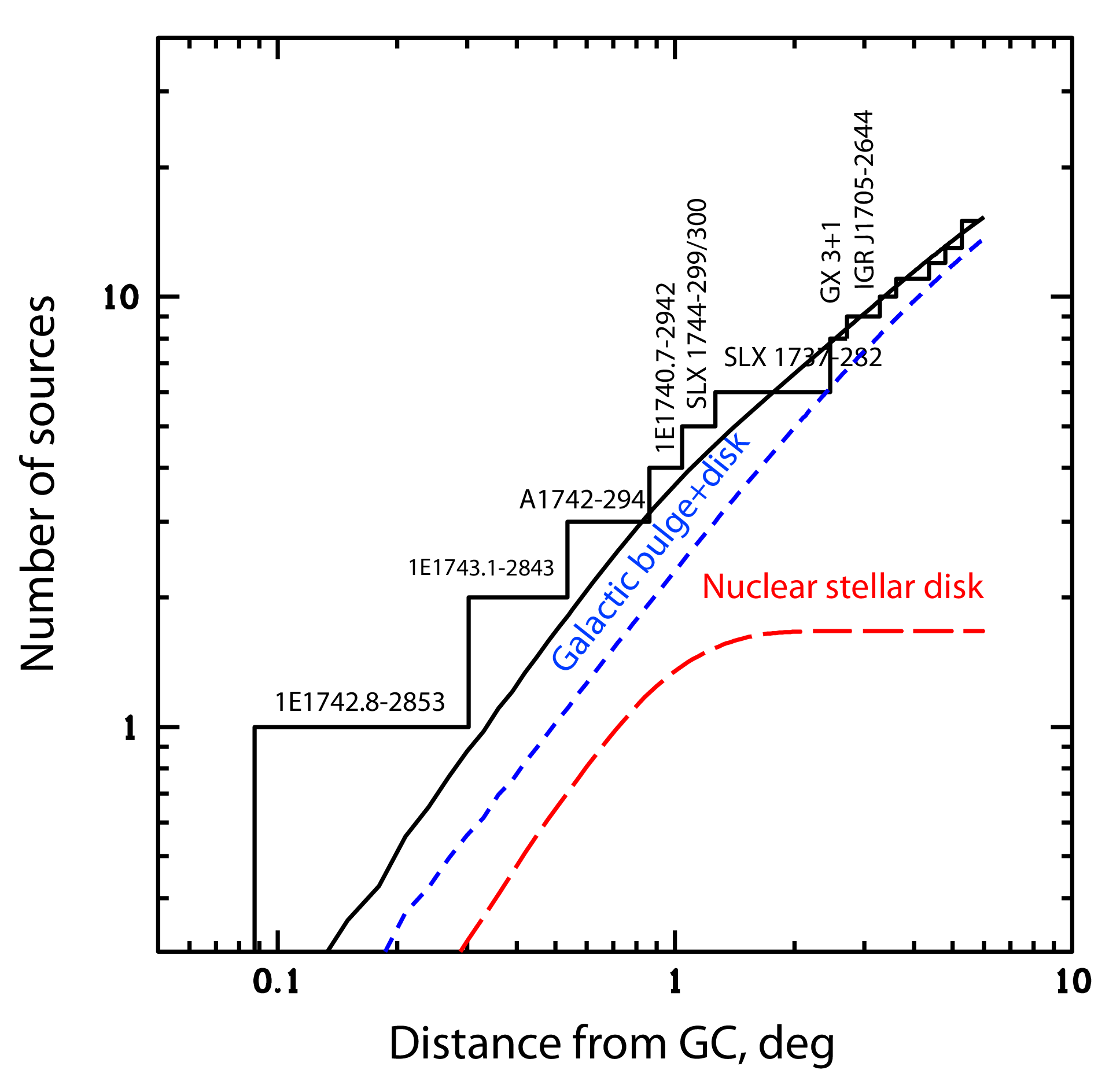}

\caption{Cumulative number of persistent LMXBs in the Milky
Way bulge seen by INTEGRAL/IBIS within the circular region of a
given projected radii. The solid curve shows the number of sources
predicted by the Galaxy mass model described above and the
LMXB LF of \cite{gilfanov04} (multiplied by 0.4). The contributions of the
Galactic bulge and nuclear stellar disk components are indicated by the
short- and long-dashed lines, respectively. For the model curve we
assumed the luminosity limit of $10^{35}$ \lum. Positions of sources
in the central region of the bulge are labeled.} \label{cumnum}

\end{figure}

An extensive study of LMXB distributions in the Milky Way and in
other galaxies showed that their number density closely traces the 
stellar mass distribution
\cite[e.g.][]{gursky75,skinner93,grebenev96,grimm02,gilfanov04}. A
rough comparison of the surface density of LMXBs in the Galactic bulge
region with that of stars was previously done by \cite{skinner93}
based on archival observations with different instruments and by
\cite{grebenev96} based on a smaller sample of sources detected with
the ART-P telescope aboard the {\em GRANAT} observatory. Now, using
the sample of sources detected by INTEGRAL over five years of its 
operation we can make a somewhat better comparison.

The cumulative angular distribution of the number of persistent LMXBs
detected by INTEGRAL in the Galactic bulge
region as a function of projected angular distance from the Galactic
Center is shown in Fig.~\ref{cumnum} by the histogramm. The solid
curve in this figure shows the cumulative stellar mass within given
projected radii. To match the number of observed sources at 
distances $<5.1^\circ$ from the Galactic Center with the galactic mass
model described above (the total stellar mass contained in this region
is $\approx1.2\times10^{10} M_\odot$) and the average LMXB LF of
\cite{gilfanov04}, the latter was multiplied by a factor of 0.4.

\begin{figure}
\includegraphics[width=\columnwidth]{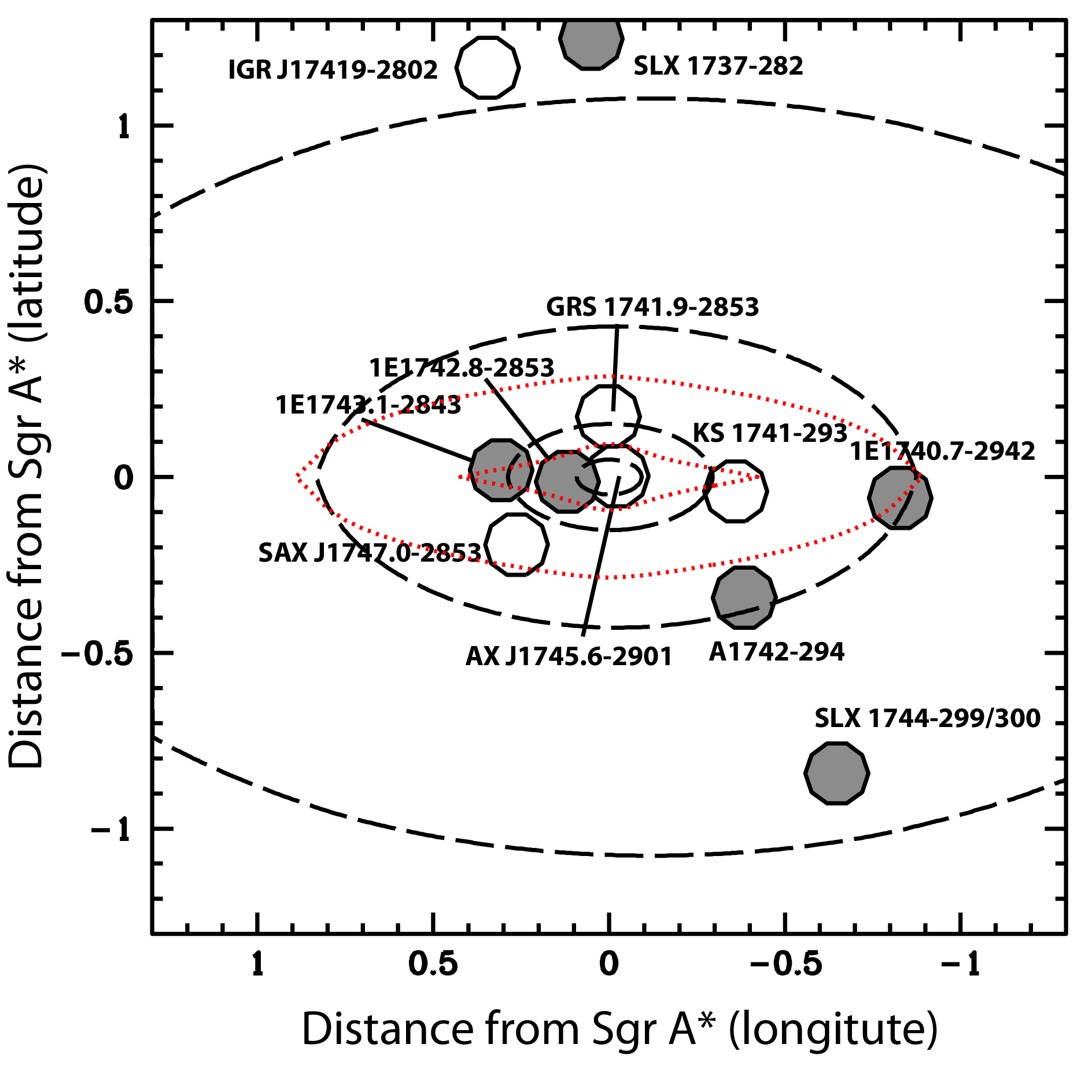}

\caption{Positions of all sources detected by IBIS/ISGRI within
the inner 1.3$^\circ$ around the center of the Galaxy (Sgr A$^*$).
Dashed (black) curves are contours of the $iso$ surface density of
the stellar mass in the model of the Galaxy, which includes the
bulge, the nuclear stellar disk and nuclear stellar cluster (which
is important only within $\sim$10~\arcmin\ around Sgr A$^*$). Dotted
curves are $iso$ surface density contours of the nuclear stellar disk
component only. Filled (gray) circles show positions of persistent
sources, open circles -- positions of transient ones.}\label{sources}

\end{figure}

Interestingly, the radial distribution of sources
shows an indication on an excess of the number of LMXBs inside the
innermost $\sim2^\circ$ with respect to that expected from the
distribution of stars in the Galactic bulge (note that a similar
effect was previousely observed by \citealt{skinner93}). The excess
itself has low statistical significance -- instead of the predicted
$\sim1$ source within the innermost $0.6^\circ$ around Sgr A$^*$ we
see three sources, but it is consistent with the presence of LMXBs
from the so-called nuclear stellar disk \citep[e.g.][]{launhardt02} (its  
contribution is shown by the long-dashed line). This allows us to 
tentatively suggest that at least the innermost sources, namely
1E1742.9$-$2853 and 1E1743$-$2843, reside in the nuclear stellar disk.

In order to visualize the distribution of sources in the innermost
region of the Galaxy (see Fig.~\ref{sources}), we present the
positions of detected LMXBs within $1.3^\circ$ around Sgr A$^*$ with
overlayed contours of $iso$ surface density of stars in the region
for the different components of the Galactic stellar population.

It is necessary to note that the best-fit LF normalization for the
Galactic bulge population of LMXBs is $\approx0.4$ of the average over
a number of nearby galaxies 
\citep{gilfanov04}. Nevertheless, the Galactic bulge normalization is
within the spread of best-fit values obtained for different galaxies
\citep{gilfanov04}. 

\subsubsection{Transients}

\begin{figure}
\includegraphics[height=\columnwidth]{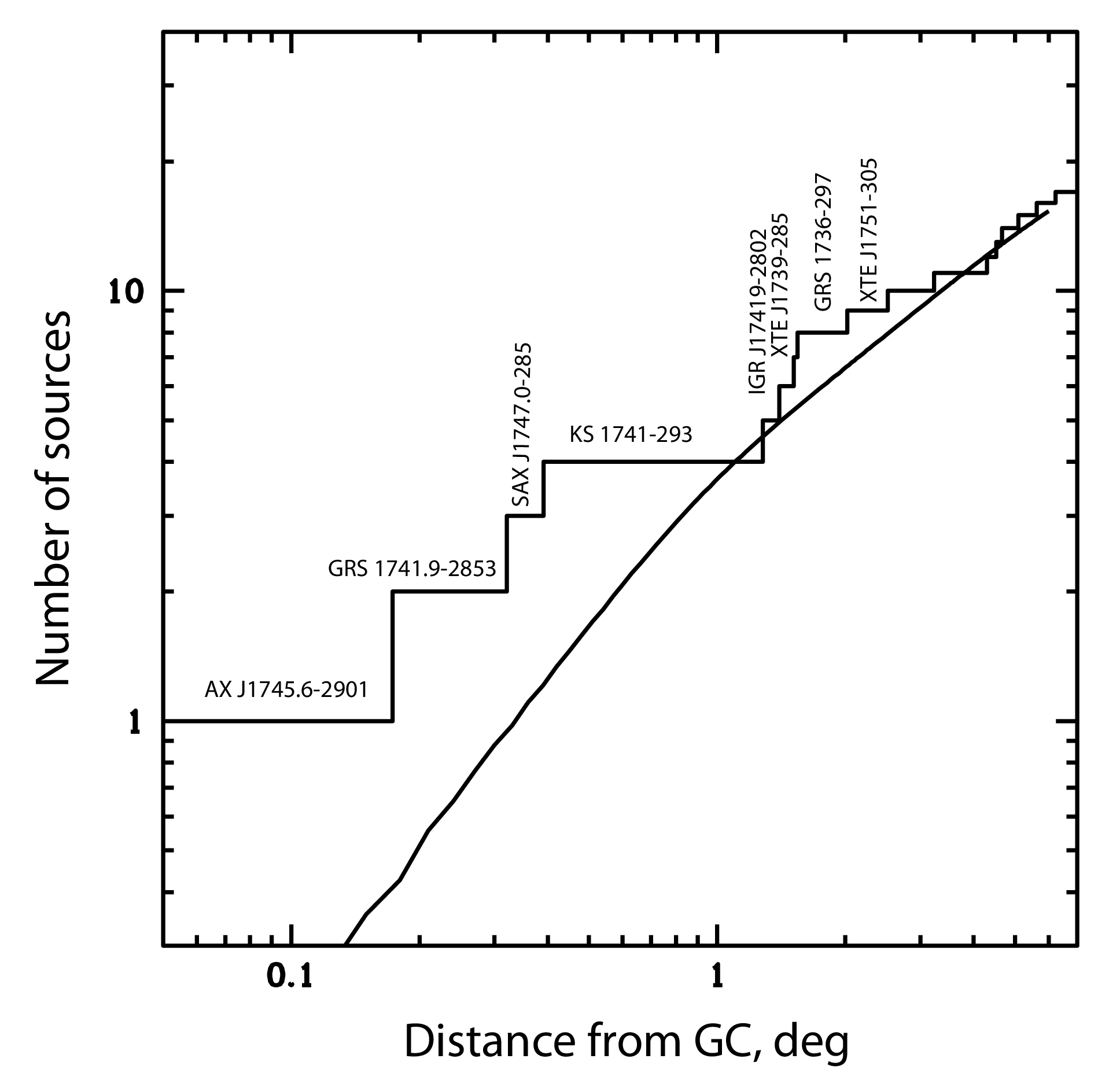}

\caption{Cumulative number of transient LMXB sources in the Galactic
Center region, as seen by INTEGRAL/IBIS. The solid line represents
the adopted mass model of the Galaxy normalized to the number of sources
within 5.1$^\circ$ of the Galactic Center. The positions of sources in
the innermost region of the bulge are labeled.}\label{cumnum_trans}

\end{figure}

Using the large time span of the INTEGRAL observations we have the
ability to effectively determine the ``transientness'' of sources
and study these transient sources separately. In particular, it is
interesting to analyze their spatial distribution in view of the
predictions of some theories that the densest central regions of 
galactic bulges might contain increased numbers of transients due
to their enhanced dynamical formation \cite[e.g.][]{voss07}.

\begin{table}
\caption{List of transients seen by INTEGRAL/IBIS over the period 2003--2007}
\label{list_transients}
\tabcolsep=3mm
\begin{tabular}{r|c|c}
\hline
\#&{Dist from Sgr A$^*$, deg}& Name\\
\hline
1&      0.02 &AX J1745.6-2901 \\
2&      0.17 &GRS 1741.9-2853\\
3&      0.33 &SAX J1747.0-2853\\
4&      0.36 &KS1 741-293\\
5&      1.27 &IGR J17419-2802\\
6&      1.36 &XTE J1739-285\\
7&      1.51 &GRS 1736-297 \\
8&      2.02 &XTE J1751-305\\
9&      2.54 &GRS 1747-313\\
10&      3.22 &IGR J17464-3213\\
11&      4.29 &SLX 1746-331\\
12&      4.48 &IGR J17353-3257\\
13&      4.67 &XTE J1807-294\\
14&      5.11 &MXB 1730-33\\
15&      5.64 &IGR J17331-2406\\
16&      6.18 &XTE J1720-318\\
17&      7.14 &A 1744-361\\
18&      7.68 &IGR J17597-2201\\
19&      7.95 &XTE J1817-330\\
20&      8.10 &4U 1746-37\\
21&      8.56 &XTE J1818-245\\
22&      8.57 &4U 1705-32\\
23&      8.68 &4U 1745-203\\
24&     10.60 &IGR J17098-3628\\
25&     10.66 &IGR J17091-3624\\
\hline
\end{tabular}
\end{table}

The list of transient sources detected by INTEGRAL within the region
of our study is presented in Table \ref{list_transients} along
with their distances from the Galactic Center. The cumulative number
of these sources as a function of their distance from the Galactic
Center is shown in Fig.~\ref{cumnum_trans}. The cumulative mass
of stars within the considered region, normalized to the number of
transients at $<5.1^\circ$ is shown by a solid line. One can note that
there is a slight indication that transient sources are more
concentrated towards the Galactic Center than the persistent ones,
but the statistical significance of this effect is low. We should note that
a similar effect was observed by \cite{skinner93}, since the majority of
the sources considered in that work were transients detected with
different instruments over a long period. If the effect of the
increased surface density of transient sources in the center of the
Galaxy is real, this might be related to the mechanism of
dymanical formation of LMXBs proposed by \cite{voss07}.

\section{Conclusions}

We analyzed the luminosity function and spatial distribution of
low-mass X-ray binaries in the central part of our Galaxy (Galactic 
bulge) using data from the INTEGRAL observatory in the 17--60 keV
energy band and found that: 

\begin{enumerate}

\item No sources in the considered region demonstrate hard X-ray (17--60 keV)
luminosities higher than $L_{\rm 17-60~keV}\sim10^{37}$ \lum. 
This happens because: a) sources with high bolometric luminosities have 
soft spectra with only a small portion of the bolometric luminosity
emitted in the hard X-ray band, b) those sources with hard spectra do
not have high bolometric luminosities. In fact, in our Galaxy there is
only one source with hard X-ray luminosity as high as $L_{\rm
17-60~keV}\sim5\times10^{37}$ \lum\ -- GRS 1915+105, which is pecular
in many respects. 

\item The luminosity function of persistent LMXBs is fairly flat at the faint
end ($L_{\rm 17-60}\la 10^{37}$ \lum): the slope is $d\log N/d\log
L=-0.96\pm0.20$. The shape of the LF of the Galactic bulge LMXBs in
the standard X-ray band is consistent with that of the LMXB LF averaged
over several nearby galaxies \citep{gilfanov04} and thus confirms the
presence of the  break at $L_{\rm x}\sim 10^{37}$ proposed earlier by
\cite{primini93} and \cite{gilfanov04}. 

\item
The origin of the observed break in the LF is not clear. One of the 
possibilities is that it is connected with a change of the mechanism
of removal of the angular momentum of the binary system: from 
magnetic stellar wind braking to graviational braking
\cite[see][]{postnov05}. We propose that some additional flattening of
the LMXB LF might be provided by the change of the mass accretion
rate--X-ray luminosity relation expected in radiatively inefficient
accretion flow models. 

\item The cumulative angular distribution of the persistent LMXBs in the
Galactic center region traces the cumulative number of stars in the
model consisting of the Galactic bulge and nuclear stellar
disk components. The data favor (though with low statistical significance) 
the presence of the nuclear stellar disk component in the distribution
of LMXBs. 

\item The cumulative angular distribution of the transient LMXBs slightly
differs from that of the persistent sources, indicating some
increase of the fraction of transient sources in the
innermost regions of the Galaxy. However, the statistical significance
of this increase is low.

\end{enumerate}

\begin{acknowledgements}

This research made use of data obtained from the High Energy
Astrophysics Science Archive Research Center Online Service,
provided by the NASA/Goddard Space Flight Center. This work was
supported by DFG-Schwerpunktprogramme (SPP 1177), grant of Russian
Foundation of Basic Research 07-02-01051, 07-02-00961-a, NSh-5579.2008.2 and
programm of Presidium of RAS ``The origin and evolution of stars and
galaxies'' (P04).

\end{acknowledgements}

\end{document}